# A Large and Precise All-Sky Photometric Standard Star Dataset Across More Than 200 Passbands

Kai Xiao[1,2,*], Yang Huang[1,3,*], Haibo Yuan[2,4,*], Bowen Huang[2,4], Dongwei Fan[5,6], Timothy C. Beers[7], Zhirui Li[3,1], Henggeng Han[3], Qiqian Zhang[3,1], Tao Wang[2,4], Mingyang Ma[2,4], Yuanchang Wang[2,4], Shuai Xu[2,4], Lin Yang[8], and Jifeng Liu[1,3,2,*]

[1]School of Astronomy and Space Science, University of Chinese Academy of Sciences, Beijing 100049, China
[2]Institute for Frontiers in Astronomy and Astrophysics, Beijing Normal University, Beijing, 102206, China
[3]National Astronomical Observatories, Chinese Academy of Sciences, Beijing 100101, China
[4]School of Physics and Astronomy, Beijing Normal University No.19, Xinjiekouwai St, Haidian District, Beijing, 100875, China
[5]Department of Physics and Astronomy, The Johns Hopkins University, 3400 North Charles Street, Baltimore, MD 21218, USA
[6]National Astronomical Data Center, Beijing 100101, China
[7]Department of Physics and Astronomy and JINA Center for the Evolution of the Elements (JINA-CEE), University of Notre Dame, Notre Dame, IN 46556, USA
[8]Department of Cyber Security, Beijing Electronic Science and Technology Institute, Beijing, 100070, China
[*]corresponding authors: xiaokai@ucas.ac.cn, huangyang@ucas.ac.cn, yuanhb@bnu.edu.cn and jfliu@nao.cas.cn

## ABSTRACT

High-precision photometric standard stars play a key role in enabling accurate photometric calibration and advancing various fields of astronomy. However, due to limitations in calibration methods and the limited availability and underuse of high-precision reference data, existing photometric standard stars may suffer from insufficient numbers, systematic errors exceeding 10 milli-magnitude (mmag), limited photometric band coverage, or incomplete sky coverage, among other issues. To overcome these limitations, we have constructed the largest (over 200 million stars, 1000 times the widely recognized Landolt standards in the same magnitude range), most precise (better than 10 mmag), and most comprehensive (over 200 bands, nearly 40 times the coverage of traditional standards) all-sky standard stars. Based on standards, we have calibrated multiple survey datasets to mmag precision, and subsequently developed a complete sky distribution of stars for the Pan-STARRS system. This database, the BEst STars Database (BEST), is expected to pave the way for achieving mmag-level - or even higher - photometric precision in large-scale surveys, and to play a central role in shaping a high-precision astronomical measurement framework.

## Background & Summary

Astronomy, as an observation-driven discipline, is critically dependent on uniform and accurate photometric calibration. The successful implementation of the Sloan Digital Sky Survey (SDSS)[1] at the turn of this century marked a major milestone, offering new insights into our understanding of the Universe and inaugurating the era of large-scale digital sky surveys. Over the past decades, numerous digital photometric surveys have steadily emerged, such as the Panoramic Survey Telescope and Rapid Response System (Pan-STARRS)[2], SMSS Southern Sky Survey (SMSS)[3] and the Dark Energy Survey (DES)[4]. The rich trove of multi-band photometric data from these surveys has greatly advanced astronomical research and modern astrophysics.

However, the photometric calibration processes of these surveys vary in terms of strategy, methodology, and achieved precision, and procedures are often complex, involving multiple assumptions and operational requirements. For example, SDSS calibration assumes that the one-dimensional flat-field remains constant within a flat-field season, and that the nightly atmospheric extinction coefficient varies linearly with time[5]. The DES calibration adopts a forward-modeling approach to account for atmospheric and instrumental effects, requiring auxiliary instruments for high-precision atmospheric monitoring, a sufficient number of nightly exposures, and multi-band repeated observations[6]. Such assumptions either limit the achievable precision or increase the calibration complexity due to operational demands. A fundamental reason underlying these constraints is the lack of sufficiently numerous, high-precision photometric standard stars. For example, although the widely used Landolt standard stars[7] offer approximately 1% mag photometric precision, they number only a few tens of thousands of stars and are mainly distributed near the celestial equator. Similarly, while the ATLAS all-sky stellar reference catalog[8] (http://dx.doi.org/10.17909/t9-2p3r-7651) provides all-sky coverage, its photometric accuracy suffers from systematic errors at the 2-3% mag level.

In recent years, the number of wide-field and time-domain photometric surveys has grown rapidly. Notable examples include the Javalambre Physics of the Accelerating Universe Astrophysical Survey (J-PAS)[9], which provides photometry in 54 narrow/medium bands and 3 medium bands; the Javalambre Photometric Local Universe Survey (J-PLUS)[10] and the Southern Photometric Local Universe Survey (S-PLUS)[11], offering photometry in 7 narrow/medium bands and 5 broad bands; and other surveys such as the Stellar Abundance and Galactic Evolution Survey (SAGES)[12], the Chinese Space Station Telescope (CSST)[13], the Legacy Survey of Space and Time (LSST)[14], the Multi-channel Photometric Survey Telescope (Mephisto)[15], and the SiTian Project (SiTian)[16]. These projects span near-ultraviolet to near-infrared wavelengths, use varied survey strategies and detectors (CCDs and CMOS), and field-of-view from a few arcminutes to hundreds of square degrees.

This diversity in instrumentation and observational design presents significant challenges, highlighting the urgent need for ultra-high-precision photometric calibration. Establishing a large number of high-precision photometric standard stars, tailored to modern, complex photometric systems and pass-bands, alongside dedicated calibration strategies, would greatly streamline the photometric calibration process and facilitate efficient, high-precision measurements.

Large-scale spectroscopic surveys such as LAMOST[17-20], combined with high-precision photometric data (e.g., Gaia), offer the opportunity to construct millions of photometric standard stars using the stellar color regression (SCR)[21] method. The corrected Gaia BP/RP (XP) spectra[22, 23], calibrated using CALSPEC[22, 24] (https://archive.stsci.edu/hlsps/reference-atlases/cdbs/calspec/), Hubble's Next Generation Spectral Library[25] (http://archive.stsci.edu/prepds/stisngsl/), and LAMOST DR7 data (https://dr7.lamost.org/v1.3/), effectively remove systematic errors as functions of color, G magnitude, and extinction, achieving consistency better than 2% in the near-ultraviolet and 1% at longer wavelengths within $-0.5 <$ BP$-$RP $< 2$, $3 <$ G $< 17.5$, and $E(B-V) < 0.8$. These corrected spectra, obtained by applying the systematic error correction relations23 to the original Gaia XP spectra[22] (https://cdn.gea.esac.esa.int/Gaia/gdr3/Spectroscopy/xp_continuous_mean_spectrum/), further enable the creation of billions of all-sky standard stars through the improved Gaia XP based synthetic photometry (XPSP) method[26]. Both methods have already demonstrated their potential for constructing high-precision photometric calibration standards, as shown by their application to previous datasets. The SCR method, which achieves 1-5 milli-magnitude (mmag) precision, has improved the calibration of SDSS Stripe 82[27] colors[21] (https://faculty.washington.edu/ivezic/sdss/catalogs/stripe82.html) and magnitudes[28], Gaia DR2[29, 30] and EDR3[31-34] (https://cdn.gea.esac.esa.int/Gaia/), SMSS DR[23, 35], SAGES[12, 36, 37] uv- and gri-bands[38], and Pan-STARRS DR1[39, 40, 41]. The XPSP method can achieve similar precision to the SCR method, as demonstrated by the recalibration of J-PLUS DR3[26] (https://archive.cefca.es/catalogues/jplus-dr3) and S-PLUS DR4[65] (https://splus.cloud/).

This paper utilizes stellar parameters from large-scale spectroscopic surveys like LAMOST, photometric data from Gaia, and corrected Gaia XP spectra, applying both the SCR and the XPSP methods to create a database of billions of photometric standard stars brighter than G = 17.65 mag, distributed over the full sky. Using this database, we then perform a mmag precision photometric calibration on datasets from photometric systems such as SMSS[3] DR4[42], resulting in a recalibrated SMSS DR4 and corrected Pan-STARRS DR1 data. Next, we establish a stellar standard star database within the Pan-STARRS system, which contains approximately one billion stars distributed over the full sky.

This database stands as the largest and most precise photometric standard to date, paving the way for both improving the precision of astronomical photometry and exploring new frontiers in astronomical high-precision investigations. For example, using the BEST database, time-domain data from the CMOS-based MST array[43], 12-band (7 narrow/medium bands and 5 broad bands) photometry from the CCD-based S-PLUS USS DR1[44] (https://doi.org/10.12149/101504), and photographic photometry (https://www.doi.org/10.12149/103032) have been calibrated to ~1 mmag[45], 1-6 mmag[46], and better than 0.1 mag[47] in their zero-points, respectively. Furthermore, Hu et al.[48] have used the BEST database to identify a sample of blue horizontal-branch stars in the Solar neighborhood with accurately determined distances, while Huang et al.[49] have derived stellar metallicities for over 100 million stars. As photometric precision continues to advance, we will gain ever clearer insight into the past, present, and future of the Universe, and into the fundamental laws that shape its evolution.

# Methods

This section presents the improved XPSP method, based on the corrected Gaia XP spectra[23], and the spectroscopy-based SCR method, both of which are used to construct standard photometric stars. The well-calibrated photometry and the all-sky distributed stellar photometry in the five Pan-STARRS bands are also described. A schematic diagram of the methodology is provided in Figure 1 to better understand the technical workflow.

## Generation of Over 200 Million All-Sky Standard Stars Across More Than 200 Photometric Pass-bands Using Corrected Gaia XP Spectra

### Data pre-processing

We downloaded the transmission functions of various photometric systems, including Gaia EDR3, SDSS, Pan-STARRS, SMSS, J-PLUS, S-PLUS, J-PAS, APASS, Hipparcos, Tycho, HST, TESS, Kepler, ZTF, LSST and CSST, from the SVO filter profile service[50-52] (http://svo2.cab.inta-csic.es/svo/theory/fps3/). These transmission functions account for the effect of filter transmission efficiency, detector quantum efficiency, and optical system effects. For the ground-based photometric systems, the Earth's atmospheric extinction is also considered (at 1.2 airmass). In particular, we obtained the five-band transmission functions for the Landolt system[53], and we thank the Mephisto and SiTian team for providing the six-band and g-band transmission functions for the Mephisto and SiTian photometric systems, respectively. All the transmission functions used in this work are presented in Figure 2.

The downloaded transmission functions were linearly interpolated onto the Gaia XP spectral scale, defined with data points every 20Å from 3360Å to 10200Å, for bands fully covered by the Gaia XP spectral wavelength. Interpolation is performed using the *interp1d* function from the SciPy package[56] in Python. For bands not fully covered, such as the SDSS u-band, we first extrapolated the spectra using linear fitting[26] and then interpolated the transmission function onto the Gaia XP spectral scale. For bands entirely outside the Gaia

XP coverage, such as the CSST NUV-band, they are not provided in the current version, but may be included in future versions as more spectral data becomes available (e.g., CSST slitless spectra) or methods improve.

### The Gaia XP spectra based synthetic photometry method

The concept of synthetic photometry was introduced as early as Cousins & Jones[54], with the core idea of mapping the stellar spectral energy distribution (SED) at the top of Earth's atmosphere to a defined photometric passband. The release of Gaia XP spectra for more than 200 million stars provided new momentum for this method. Montegriffo et al.[55] subsequently developed the Gaia DR3 XP spectra based synthetic photometry (XPSP) method, enabling the construction of more than 200 million standard stars. However, Gaia DR3 XP spectra contain magnitude- and color-dependent systematic errors[23, 55], as well as extinction-dependent systematics[23]. These introduce corresponding systematics in the XPSP standard stars - particularly in the blue bands - where magnitude errors can exceed 0.01 mag. Fortunately, Huang et al.[23] carefully corrected the Gaia DR3 XP spectra, and based on these corrected spectra, Xiao et al.[41] improved the XPSP method to improve the precision of standard stars. The improved XPSP method has since been validated by several studies[26, 41, 65].

Following the work of Xiao et al.[41], we derive the magnitudes by convolving the corrected Gaia XP spectra flux $f_\lambda(\lambda)$ with the transmission function $S(\lambda)$ for the $\chi$-band.

$$m_{\mathrm{XPSP}}^{\chi} = -2.5 \cdot \log_{10} \frac{\int f_\lambda(\lambda) \times S_\chi(\lambda) \times \lambda \times \mathrm{d}\lambda}{\int S_\chi(\lambda) \times (c/\lambda) \times \mathrm{d}\lambda} - 48.60 \quad (1)$$

Here, $\lambda$ represents the wavelength in Angstroms, and $c$ is the speed of light ($3\times10^{18}$ Å/sec). Prior to integration, we converted the flux of Gaia XP spectra from SI to cgs units by multiplying by 100. Integration is performed using the *integrate* function from the SciPy package[56] in Python. The resulting XPSP synthetic photometry is expressed in the AB magnitude system[57].

We calculate the error in the synthetic magnitude for the $\chi$-band, using the propagated spectral error.

$$\sigma_{\mathrm{XPSP}}^{\chi} = \left|\frac{-2.5}{\ln 10}\right| \times \frac{\sqrt{\sum_\chi \sigma_\lambda(\lambda)^2 \times S_\chi(\lambda)^2 \times \lambda^2 \times \Delta\lambda^2}}{\int f_\lambda(\lambda) \times S_\chi(\lambda) \times \lambda \times \mathrm{d}\lambda} \quad (2)$$

Here, $\sigma_\lambda$ represents the flux error of Gaia XP spectra, and $\Delta\lambda$ is 20Å. The unit conversion and integration process are the same as when calculating magnitudes.

## Construction of SCR Standard Stars Using Multi-Spectroscopic Stellar Atmospheric Parameters

### Data pre-processing

The spectroscopic data from the LAMOST[17-20] DR11 (https://www.lamost.org/dr11/) and the Galactic Archaeol-ogy with HERMES[58] (GALAH) DR4 (https://www.galah-survey.org/dr4/) are downloaded and used. Subse-quent updates and additional results will be provided on the website. We also acquired Gaia EDR3 photometric data[22].

We first applied a magnitude-dependent correction[34] to the Gaia EDR3 photometry (hereafter Gaia EDR3 photometry), and combined the spectroscopic data from the LAMOST DR11[17-20] and GALAH DR4[58] with the high-precision photometric data from Gaia EDR3 using a cross-matching radius of 1". Main-sequence stars were then selected using Equation (1)[49] with the following constraints:

- For Gaia EDR3 photometry, `phot_bp_rp_excess_factor` $< 1.3 + 0.06 \times (G_{BP} - G_{RP})^2$ was used to exclude poor-quality Gaia $G_{BP}$ or $G_{RP}$ photometry;

- For LAMOST DR11 and GALAH DR4, stars with $4500 < T_{\rm eff} < 6500$K and $T_{\rm [Fe/H]} > -1$ were selected for better intrinsic color fitting with stellar atmospheric parameters, and a signal-to-noise ratio of at least 20 for the g-band (SNRG) and SNR_c2_iraf, respectively. For GALAH DR4, flag_sp ≤ 1 was used;

- The reddening values $E(B-V)_{\rm SFD}$ were obtained from the 2D reddening map[59] for the SCR sample stars.

## Estimating the extinction value of E($G_{BP}$−$G_{RP}$) with the star-pair method

We avoid using the dust reddening map by SFD[59] for constructing the SCR photometric standard stars due to its unreliability at low Galactic latitudes and spatially-dependent systematic errors[60]. In this section, the values of $E(G_{BP}-G_{RP})$ obtained with the star-pair method[61, 62] are predicted and adopted instead.

The core idea of the star-pair method is that stars with the same atmospheric parameters share the same intrinsic colors. The calculation process is as follows:

- First, we select reference stars from the SCR standard stars based on following constraints: Galactic latitude higher than 30°, $E(B-V)_{\rm SFD}$ less than 0.02 mag, and distances from the Galactic plane greater than 300 pc. These constraints help balance the reduction of systematic errors in $E(B-V)_{\rm SFD}$ with ensuring that reference stars are evenly distributed in parameter space and are sufficiently numerous.

- Then, the color $E(G_{BP} - G_{RP})$ of the SCR sample stars is adopted using Gaia EDR3 photometry, and the intrinsic color $(G_{BP}-G_{RP})_0$ is determined as follows:

$$(G_{\rm BP} - G_{\rm RP})_0 = G_{\rm BP} - G_{\rm RP} - E(G_{\rm BP} - G_{\rm RP}) \quad (3)$$

- To solve for $E(G_{BP}-G_{RP})$, we employ a numerical iteration method. First, an initial value for $E(G_{BP}-G_{RP})$ is provided, derived from $R(G_{BP}-G_{RP}) \cdot E(B-V)_{\rm SFD}$, and the intrinsic color $(G_{BP}-G_{RP})_0$ is predicted using Equation (3). Here, $R(G_{BP}-G_{RP})$ is a second-order two-dimensional polynomial, as a function of $E(B-V)_{\rm SFD}$ and $T_{\rm eff}$, with six coefficients[33].

For each star in the SCR sample (with $T_{\rm eff, i}$, [Fe/H]$_i$ and log $g_i$), reference stars are selected based on the following conditions: $|T_{\rm eff} - T_{\rm eff, i}| < T_{\rm eff, i} \times (3 \times 10^{-5} \times T_{\rm eff, i})^2$ K, $|\log g - \log g_i| < 0.5$ dex, and $|[{\rm Fe/H}] - [{\rm Fe/H}]_i| < 0.3$ dex[62]. Next, for reference stars, a first-order three-dimensional polynomial (with four free parameters: $a_0$, $a_1$, $a_2$, $a_3$, see Equation (4)) is used to fit the intrinsic colors, as a function of $T_{\rm eff}$, [Fe/H], and log $g$. This polynomial is then applied to determine the intrinsic color for each SCR sample star using Equation (4). The predicted intrinsic color is substituted back into Equation (1) to obtain $E(G_{BP}-G_{RP})$, and the process is iterated.

$$(G_{\rm BP}-G_{\rm RP})_0 = a_0 \cdot T_{\rm eff} + a_1 \cdot [{\rm Fe/H}] + a_2 \cdot \log g + a_3 \quad (4)$$

## The SCR method

The SCR method, first introduced by Haibo Yuan[21], centers on the idea of predicting the intrinsic colors of stars using a few physical quantities. The intrinsic colors can be estimated from spectroscopic data, either by using atmospheric parameters derived from stellar spectra (the spectroscopy-based SCR method[21]) or directly from the spectra. Alternatively, photometric data can be used to predict intrinsic colors from multi-band colors[34] or stellar loci based on photometric metallicity (the photometry-based SCR method[41]). Only the spectroscopy-based SCR method is applied to construct photometric standard stars in the current BEST database version, but future versions may incorporate other methods.

The SCR method used here involves four steps:

- First, for each χ-band, a color $G_{BP/RP} - m_{XPSP}^{\chi}$ is adopted from Gaia photometry and the XPSP standards. The intrinsic color $(G_{BP/RP} - m_{XPSP}^{\chi})_0$ is expressed as:

$$(G_{BP/RP} - m_{XPSP}^{\chi})_0 = G_{BP/RP} - m_{XPSP}^{\chi} - R_{(G_{BP/RP} - m_{XPSP}^{\chi})} \times E(G_{BP} - G_{RP}) \quad (5)$$

where $G_{BP/RP}$ refers to $G_{BP}$ for photometric bands with central wavelengths below 520 nm, and $G_{RP}$ for those above 520 nm.

- Secondly, using low-extinction stars with $E(G_{BP} - G_{RP}) < 0.01$, the relationship $\mathbb{F}(T_{eff}, [Fe/H])$ between intrinsic colors $(G_{BP/RP} - m_{XPSP}^{\chi})_0$ and stellar atmosphere parameters ($T_{eff}$ and $[Fe/H]$) is defined by second/third-order two-dimensional polynomial fitting, based on an initial constant value[63] of $R_{(G_{BP/RP} - m_{XPSP}^{\chi})}$. Polynomial fitting in this work is performed using the *curve_fit* function from the SciPy package[56] in Python.

- Thirdly, the relationship is used to derive the intrinsic colors for all of the SCR sample stars, and a second-order two-dimensional polynomial, $\mathbb{R}(T_{eff}, E(G_{BP} - G_{RP}))$, as a function of $T_{eff}$ and $E(G_{BP} - G_{RP})$, is then used to fit $R_{(G_{BP/RP} - m_{XPSP}^{\chi})}$, estimated by:

$$R_{(G_{BP/RP} - m_{XPSP}^{\chi})} = \frac{G_{BP/RP} - m_{XPSP}^{\chi} - \mathbb{F}(T_{eff}, [Fe/H])}{E(G_{BP} - G_{RP})} \quad (6)$$

Substituting $\mathbb{R}(T_{eff}, E(G_{BP} - G_{RP}))$ into the second step, replacing the initial value and iterating, produces the true reddening coefficient $\mathbb{R}(T_{eff}, E(G_{BP} - G_{RP}))$, and the final relationship $\mathbb{F}(T_{eff}, [Fe/H])$. Statistical uncertainties for different bands in the SCR method, quantified by the standard deviation of residuals from a polynomial fit, are provided in the Technical Validation section.

- Finally, after a precise reddening correction, the SCR standard magnitude for the χ-band is determined as:

$$m_{SCR}^{\chi} = G_{BP/RP} - \mathbb{F}(T_{eff}, [Fe/H]) - \mathbb{R}(T_{eff}, E(G_{BP} - G_{RP})) \times E(G_{BP} - G_{RP}) \quad (7)$$

## Recalibration of Gaia DR3, Pan-STARRS DR1, SMSS DR4, J-PLUS DR3, S-PLUS DR4, and USS DR1 Photometry to a Few mmag Precision

For photometric data that have undergone high-precision calibration, such as Gaia EDR3, Pan-STARRS DR1, J-PLUS DR3, S-PLUS DR4 and USS DR1, we apply the systematics correction relationship to all survey data to address systematic errors. The corrected photometric data are available on the website. Please note that the BEST database contains far more than just these photometric data. The current version includes only this subset, with additional data to be released on the website in the future. The detailed description is as follows:

- An independent validation[34] of Gaia DR3 photometry was conducted using 10,000 Landolt standard stars. A machine-learning technique converted Landolt UBVRI magnitudes to Gaia magnitudes and colors, which were then compared with those in EDR3, accounting for metallicity effects. The results confirm significant improvements in the Gaia EDR3 calibration. However, modest trends up to 10 mmag were found in the G-band magnitudes ($10 < G < 19$), especially at the bright and faint ends. Using synthetic magnitudes from CALSPEC spectra with Gaia EDR3 pass-bands, absolute corrections were determined, optimizing Gaia EDR3 photometry for high-precision studies. In the BEST database, photometric corrections for all stars were applied using magnitude error correction data[34].

• For Pan-STARRS DR1, independent photometric validation and calibration was performed using LAMOST DR7 and corrected Gaia EDR3 data using the SCR method[40]. The precision of Pan-STARRS in the grizy filters is about 5-7 mmag, with significant spatial variations in magnitude offsets (up to 0.01 mag) due to calibration errors. Magnitude-dependent errors (0.005 mag for griz and 0.003 mag for y) were also found, likely from PSF magnitude uncertainties. Larger errors (up to 0.04 mag) occurred in crowded fields near the Galactic plane. Recalibration using the SCR and XPSP methods confirmed these variations, with consistency within 1-2 mmag. Two-dimensional correction maps and a Python package (https://doi.org/10.12149/101283) are available for high-precision studies and as references for other surveys[41]. In the BEST database, photometric corrections for all stars were applied using the results of the aforementioned works[40, 41].

• For J-PLUS DR3, S-PLUS DR4, and USS DR1, photometry was validated and recalibrated using corrected Gaia EDR3 data, LAMOST DR7 spectroscopy, and tens of thousands of XPSP and SCR standard stars. For J-PLUS DR3, systematic errors up to 15 mmag were found from the SL method and 10 mmag from Gaia XP spectra, with the improved XPSP method achieving 1-5 mmag agreement with SCR, improving the zero-point precision by a factor of two[26]. S-PLUS DR4 photometry exhibited position-dependent errors up to 23 mmag, with 1-6 mmag offsets between XPSP and SCR, correcting CCD position-dependent errors[64]. For USS DR1, we identified spatial variations in zero-point offsets up to 30-40 mmag for blue filters and 10 mmag for others, likely from calibration errors. Large CCD position-dependent errors (up to 50 mmag) were corrected using a second-order two-dimensional polynomial fitting and stellar flat-field corrections. Recalibrated results from XPSP and SCR standards were consistent within 6 mmag, improving zero-point accuracy sixfold. Further validation with SDSS Stripe 82[28], Pan-STARRS[40, 41], LAMOST, and Gaia photometry confirms this precision.

• For SMSS DR4 photometry, we performed an independent validation and recalibration using photometric standard stars in the SMSS photometric system, developed using the XPSP and the SCR methods discussed in the previous section. Initially, we detected spatial variations in zero-point differences between the XPSP standard stars and the SMSS DR4 data, which were attributed to systematic errors in the SMSS data. To correct these errors, we generated a two-dimensional spatial correction map with a spatial resolution of 14'.

# Prediction of Southern-Sky Standard Stars in the Pan-STARRS Photometric System from SMSS DR4 Photometry

In this section, we describe the process for constructing all-sky standard stars in the Pan-STARRS Photometric System ($g_{\text{Pan-STARRS}}$, $r_{\text{Pan-STARRS}}$, $i_{\text{Pan-STARRS}}$, $z_{\text{Pan-STARRS}}$, $y_{\text{Pan-STARRS}}$), based on the Pan-STARRS DR1 and SMSS DR4 photometry. Other photometric systems may be included in future versions.

To accomplish this, we first combine Pan-STARRS XPSP stars with SMSS DR4 and Gaia DR3 photometry. We then select reference stars based on the following criteria: photometric errors in $g_{\text{Pan-STARRS}} r_{\text{Pan-STARRS}} i_{\text{Pan-STARRS}} z_{\text{Pan-STARRS}} y_{\text{Pan-STARRS}}$ smaller than 0.01 mag, the relative error of Gaia parallax less than 20%, `phot_bp_rp_excess_factor` $< 1.3 + 0.06 \times (G_{\text{BP}} - G_{\text{RP}})^2$, and $E(B-V)$ less than 0.02 mag. Here, $E(B-V)$ from the 3D dust map of Wang et al. (2025, in press), and the distance estimate is taken from Gaia EDR3.

Secondly, for the reference stars of the $\chi$-band, we use a third- or fourth-order one-dimensional polynomial, as a function of the intrinsic color $(G_{\text{BP}} - G_{\text{RP}})_0$, to fit the intrinsic color $(\chi_{\text{SMSS}} - \chi_{\text{Pan-STARRS}})_0$. The intrinsic colors are estimated as follows:

$$\begin{pmatrix} G_{\text{BP}} - G_{\text{RP}} \\ g_{\text{SMSS}} - g_{\text{Pan-STARRS}} \\ r_{\text{SMSS}} - r_{\text{Pan-STARRS}} \\ i_{\text{SMSS}} - i_{\text{Pan-STARRS}} \\ z_{\text{SMSS}} - z_{\text{Pan-STARRS}} \\ z_{\text{SMSS}} - y_{\text{Pan-STARRS}} \end{pmatrix}_0 = \begin{pmatrix} G_{\text{BP}} - G_{\text{RP}} \\ g_{\text{SMSS}} - g_{\text{Pan-STARRS}} \\ r_{\text{SMSS}} - r_{\text{Pan-STARRS}} \\ i_{\text{SMSS}} - i_{\text{Pan-STARRS}} \\ z_{\text{SMSS}} - z_{\text{Pan-STARRS}} \\ z_{\text{SMSS}} - y_{\text{Pan-STARRS}} \end{pmatrix} - \begin{pmatrix} 1.480 \\ 0.188 \\ 0.329 \\ 0.165 \\ 0.099 \\ 0.351 \end{pmatrix} \times E(B-V) \quad (8)$$

where $1.250^{35}$, $0.188^{35, 63}$, $0.329^{35, 63}$, $0.165^{35, 63}$, $0.099^{35, 63}$ and $0.351^{35, 63}$ are the reddening coefficients with respect to $E(B-V)$ for $G_{\text{BP}} - G_{\text{RP}}$, $g_{\text{SMSS}} - g_{\text{Pan-STARRS}}$, $r_{\text{SMSS}} - r_{\text{Pan-STARRS}}$, $i_{\text{SMSS}} - i_{\text{Pan-STARRS}}$, $z_{\text{SMSS}} - z_{\text{Pan-STARRS}}$, and $z_{\text{SMSS}} - y_{\text{Pan-STARRS}}$ colors, respectively.

Finally, we use the relationship obtained to determine the magnitudes of all stars with $(G_{\text{BP}} - G_{\text{RP}})_0$ colors as follows:

$$\begin{pmatrix} g_{\text{Pan-STARRS}} \\ r_{\text{Pan-STARRS}} \\ i_{\text{Pan-STARRS}} \\ z_{\text{Pan-STARRS}} \\ y_{\text{Pan-STARRS}} \end{pmatrix} = \begin{pmatrix} g_{\text{SMSS}} \\ r_{\text{SMSS}} \\ i_{\text{SMSS}} \\ z_{\text{SMSS}} \\ z_{\text{SMSS}} \end{pmatrix} - \begin{pmatrix} f_g \\ f_r \\ f_i \\ f_z \\ f_y \end{pmatrix} - \begin{pmatrix} 0.188 \\ 0.329 \\ 0.165 \\ 0.099 \\ 0.351 \end{pmatrix} \times E(B-V) \quad (9)$$

We note that the above process was carried out separately for dwarfs and giants, with further details available on the website (https://nadc.china-vo.org/data/best/sysadmin/cms/article/view?id=Pan-STARRSall&preview=True).

## Data Records

The dataset[64] (v1.0; https://doi.org/10.12149/101651) comprises two main directories: std and well_cali, with data volumes of approximately 810 GB and 3.2 TB, respectively. All catalogs are provided in either FITS or CSV format. The std directory contains 20 subdirectories and three FITS files. The subdirectories collectively include 219,197,643 Gaia DR3 XP-based standard stars in 20 photometric systems. The file GaiaXP_PhotInfo_replenish.fits provides Gaia information for all 219,197,643 stars (as summarized in Table 1). The other two FITS files contain standard stars derived from the SCR method based on GALAH and ALMOST. The well_cali directory includes seven datasets scaled to the mmag level. Although the number of stars and column names may vary slightly between datasets, they are largely consistent with the officially released versions. For example, the column structure of the Pan-STARRS dataset is illustrated in Table 2.

## Technical Validation

The SCR and XPSP methods used in this study are two independent approaches. Their agreement is taken as an indicator of accuracy, providing a reasonable estimate in the absence of an external reference standard. From this comparison, the precision for individual stars is estimated to be approximately 0.01-0.02 mag in the u/v-like bands and 0.001-0.005 mag in the grizy-like bands.

Furthermore, the BEST database[64] (v1.0) has been applied to multiple ground-based photometric survey datasets, consistently achieving mmag-level zero-point precision for digital surveys. For example, Xiao et al. calibrated the mini-SiTian data using the SiTian standard stars, achieving a zero-point precision of 1 mmag[45]. Li et al. used S-PLUS standard stars to recalibrate S-PLUS USS DR1 photometry (https://doi.org/10.12149/101504), improving the zero-point precision by a factor of about 6[46]. Ma et al. used the JKC standard stars to recalibrate the digitization of astronomical photographic plates of China (https://nadc.china-vo.org/res/r103032/), reaching a photometric precision better than 10% mag[47]. In addition,

Huang et al. are using the JKC standard stars to validate and recalibrate the Landolt standard stars, identifying zero-point errors of about 1-5% in the UB-bands (B. Huang et al., in press). Xiao et al. are also using the SAGES standard stars to perform photometric calibration of the SAGES DDO510-band photometric data, achieving a zero-point precision of 1 mmag (K. Xiao et al., in preparation). These efforts demonstrate the robustness and effectiveness of the constructed BEST database in this work.

## Usage Notes

In addition to the BEST dataset available at the China-VO PaperData Repository[64] (https://doi.org/10.12149/101651), we have also produced a database version (website), available at https://nadc.china-vo.org/data/best/.

## Code Availability

To construct the BEST dataset, we wrote several custom Python scripts. For the SCR method, the core scripts involve numerical interpolation and polynomial fitting. For the XPSP method, we utilized the tool available at https://nadc.china-vo.org/data/best/jobs/ (which is accessible and usable only when logged in). For the recalibration of the data and the generation of the final catalog, our core scripts include not only numerical interpolation and polynomial fitting but also the reading and writing of FITS files. The scripts related to numerical interpolation, polynomial fitting, and the processing of FITS table files can be found at https://casdc.china-vo.org/paperdata/101651/scripts/.

## Data Availability

The BEST datasets[64] (v1.0) with the core scripts are available here: https://doi.org/10.12149/101651.

# Acknowledgements

This work is supported by the National Natural Science Foundation of China grants No. 12403024, 12422303, 12222301, 12173007, 12273077 and 124B2055; the National Key R&D Program of China (grants Nos. 2023YFA1608300, SQ2024YFA160006901, 2024YFA1611601, and 2022YFF0711500); the Postdoctoral Fellowship Program of CPSF under Grant Number GZB20240731; the Young Data Scientist Project of the National Astronomical Data Center; and the China Postdoctoral Science Foundation No. 2023M743447. T.C.B. acknowledges partial support from grants PHY 14-30152; Physics Frontier Center/JINA Center for the Evolution of the Elements (JINA-CEE), and OISE-1927130; The International Research Network for Nuclear Astrophysics (IReNA), awarded by the US National Science Foundation, and DE-SC0023128; the Center for Nuclear Astrophysics Across Messengers (CeNAM), awarded by the U.S. Department of Energy, Office of Science, Office of Nuclear Physics.

# Author contributions statement

K.X., Y.H., H.B.Y., and J.F.L. designed and directed the project. K.X. completed the database construction, including data preparation, the improvement and application of various methods, and the design of strategies, and wrote the manuscript. D.W.F. developed the database structure and a website, implemented the database search functionality, and designed the user interface. B.W.H. provided guidance on using the corrected Gaia XP spectra and, together with M.Y.M., Y.C.W., S.X., and L.Y., assessed the reliability of the database. Z.R.L. enhanced the flowchart design for the manuscript and website. H.H.G. and Q.Q.Z. contributed to this work by downloading and preparing part of the data, and T.W. provided guidance on the application of the 3D extinction map. Y.H., H.B.Y., T.C.B., and J.F.L. revised the manuscript. All authors reviewed and approved the final version.

# Competing interests

The authors declare no competing interests.

# Figures

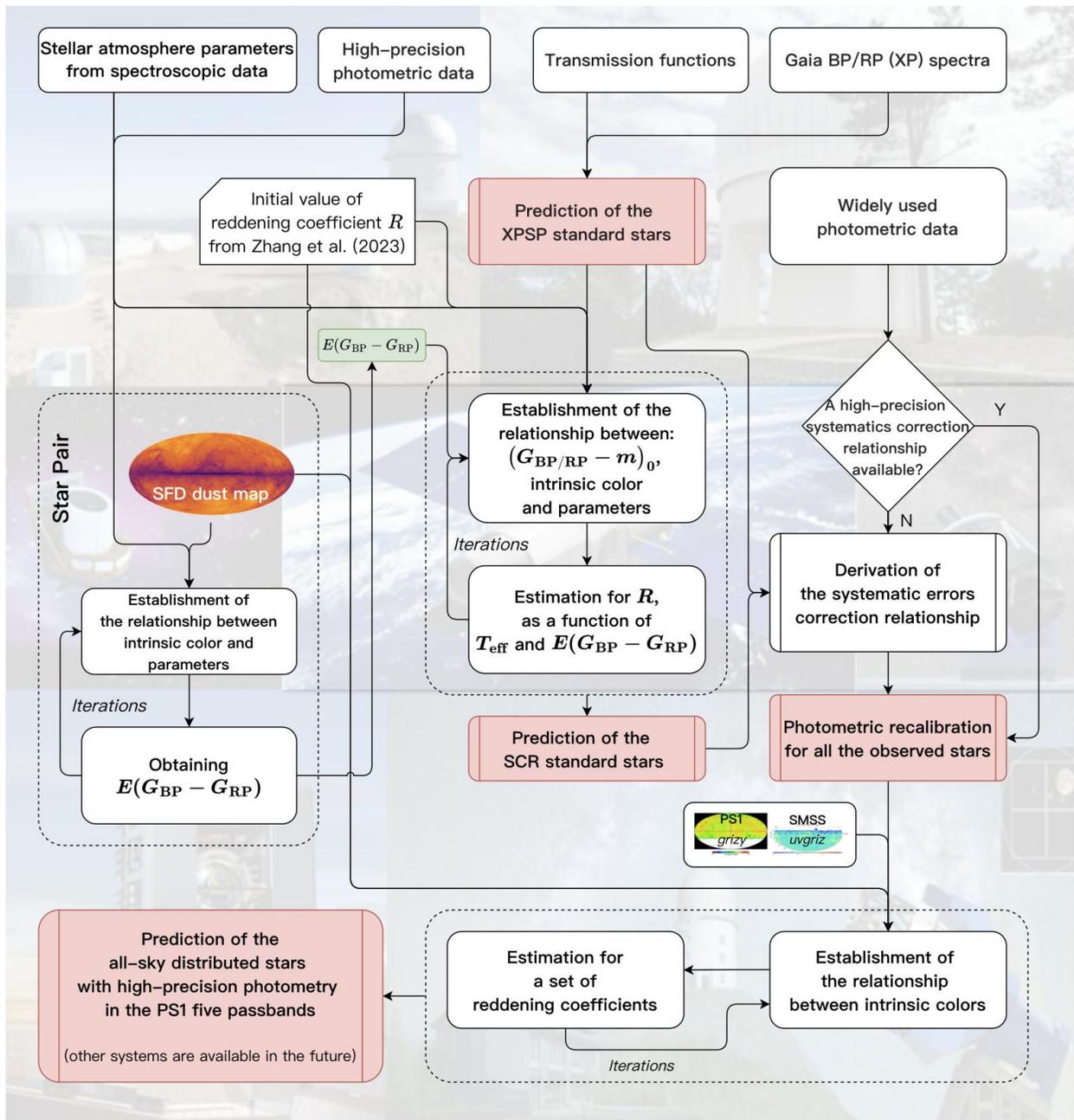

Figure 1. Flowchart of the methods in this work.

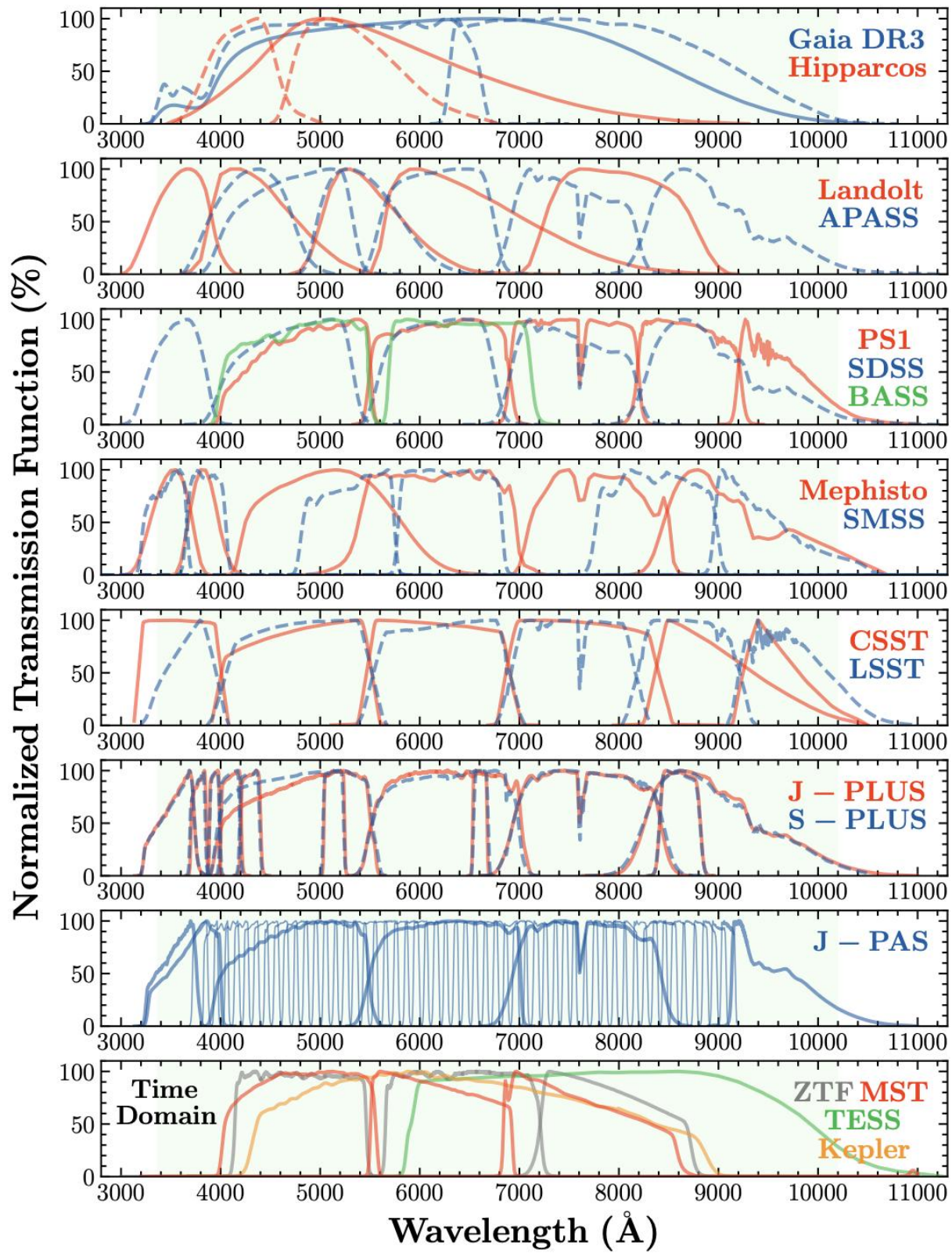

Figure 2. An example showing the transmission function of 143 filters from 18 photometric systems. The light-green shaded region in each panel indicates the wavelength coverage of the Gaia DR3 XP spectra.

# Tables

| Name | Description | Unit |
|---|---|---|
| source_id | Unique source identifier for EDR3 | - |
| l | Galactic longitude | deg |
| b | Galactic latitude | deg |
| ra | Right ascension of Gaia EDR3 in equinox J2000 | deg |
| dec | Declination of Gaia EDR3 in equinox J2000 | deg |
| pmra | Proper motion in right ascension direction | mas/year |
| pmra_error | Standard error of proper motion in right ascension direction | mas/year |
| pmdec | Proper motion in declination direction | mas/year |
| pmdec_error | Standard error of proper motion in declination direction | mas/year |
| ruwe | Renormalised unit weight error | - |
| parallax | Parallax | mas |
| parallax_error | Standard error of parallax | mas |
| phot_g_mean_mag_corr | Corrected $G$-band mean magnitude[34] | mag |
| phot_bp_mean_mag_corr | Corrected $G_{BP}$-band mean magnitude[34] | mag |
| phot_rp_mean_mag_corr | Corrected $G_{RP}$-band mean magnitude[34] | mag |
| phot_bp_rp_excess_factor | BP/RP excess factor | - |
| ebv | The 2D extinction map $E(B-V)_{SFD}$[59] | mag |
| ebprp | $E(G_{BP}-G_{RP})$ value predicted with the star-pair method[61, 62] | mag |
| x_sys_xpsp | x band magnitude of with the improved XPSP method | mag |
| x_sys_xpsp_err | x band magnitude of error with the improved XPSP method | mag |

Table 1. Description of the "corrected" Gaia DR3 XP spectra based standard stars. 'sys' represents the photometric system.

| Name | Description | Unit |
|---|---|---|
| raMean | Right ascension from single epoch detections (weighted mean) in equinox J2000 | deg |
| decMean | Declination from single epoch detections (weighted mean) in equinox J2000 | deg |
| gMeanPSFMagCorr | Corrected mean PSF magnitude from g filter detections | mag |
| gMeanPSFMagErr | Error in mean PSF magnitude from g filter detections | mag |
| gFlags | Information flag bitmask for mean object from g filter detections | - |
| rMeanPSFMagCorr | Corrected mean PSF magnitude from r filter detections | mag |
| rMeanPSFMagErr | Error in mean PSF magnitude from r filter detections | mag |
| rFlags | Information flag bitmask for mean object from r filter detections | - |
| iMeanPSFMagCorr | Corrected mean PSF magnitude from i filter detections | mag |
| iMeanPSFMagErr | Error in mean PSF magnitude from i filter detections | mag |
| iFlags | nformation flag bitmask for mean object from i filter detections | - |
| zMeanPSFMagCorr | Corrected mean PSF magnitude from z filter detections | mag |
| zMeanPSFMagErr | Error in mean PSF magnitude from z filter detections | mag |
| zFlags | Information flag bitmask for mean object from z filter detections | - |
| yMeanPSFMagCorr | Corrected mean PSF magnitude from y filter detections | mag |
| yMeanPSFMagErr | Error in mean PSF magnitude from y filter detections | mag |
| yFlags | Information flag bitmask for mean object from y filter detections | - |

Table 2. A description of the well-calibrated Pan-STARRS five-band photometry database.

| Abbreviation | Full name / Description |
| --- | --- |
| APASS | AAVSO Photometric All-Sky Survey |
| ATLAS | Asteroid Terrestrial-impact Last Alert System |
| BEST | BEst STar Database |
| CCD | Charge Coupled Device |
| CMOS | Complementary Metal-Oxide-Semiconductor |
| CSST | Chinese Space Station Telescope |
| DES | Dark Energy Survey |
| Gaia DR3 | Gaia Data Release 3 |
| GALAH | Galactic Archaeology with HERMES |
| HST | Hubble Space Telescope |
| LSST | Legacy Survey of Space and Time |
| Mephisto | Multi-channel Photometric Survey Telescope |
| MST | Mini-SiTian Array |
| NUV | Near Ultraviolet |
| TESS | Transiting Exoplanet Survey Satellite |
| J-PAS | Javalambre Physics of the Accelerating Universe Astrophysical Survey |
| J/S-PLUS | Javalambre/Southern Photometric Local Universe Survey |
| LAMOST | Large Sky Area Multi-Object Fiber Spectroscopic Telescope |
| mmag | milli-magnitude |
| Pan-STARRS | Panoramic Survey Telescope and Rapid Response System |
| SAGES | Stellar Abundance and Galactic Evolution Survey |
| SCR | Stellar Color Regression |
| SDSS | Sloan Digital Sky Survey |
| SED | Spectral Energy Distribution |
| SFD | Schlegel, Finkbeiner, and Davis 2D dust map |
| SiTian | SiTian Project |
| SMSS | SkyMapper Southern Sky Survey |
| SVO | Spanish Virtual Observatory |
| XPSP | Gaia XP based Synthetic Photometry |
| ZTF | Zwicky Transient Facility |